\documentclass{PoS}

\def\beqa{\begin{eqnarray*}}
\def\eeqa{\end{eqnarray*}}

\title{NNLL resummation for QCD cross sections}

\ShortTitle{NNLL resummation for QCD cross sections}

\author{\speaker{Nikolaos Kidonakis}%
         \thanks{This work was supported by the National Science Foundation under Grant No. PHY 0855421.}\\
        Kennesaw State University, USA\\
        E-mail: \email{nkidonak@kennesaw.edu}}

\abstract{I present results for the resummation of soft-gluon contributions 
to QCD hard-scattering cross sections at next-to-next-to-leading logarithm 
accuracy. A key ingredient is the calculation of two-loop soft anomalous 
dimensions for the partonic processes. Explicit expressions and applications 
are provided for processes that involve massless partons and/or massive quarks.
}

\FullConference{35th International Conference of High Energy Physics\\
                 July 22-28, 2010\\
                 Paris, France}

\begin{document}

\section{Resummation}

QCD corrections can be quite large for many processes in hadron colliders. 
Certain regions of kinematical phase space can contribute the dominant part 
to these corrections and can be significant even beyond NLO. One such region 
is partonic threshold, where soft-gluon corrections are important. These 
corrections are needed for increased accuracy in theoretical predictions 
and, using factorization theorems and renormalization group evolution, 
they can be resummed in terms of soft anomalous dimensions. 
Two-loop eikonal calculations of these quantities 
allow the resummation to be performed at 
next-to-next-to-leading logarithm (NNLL) accuracy.

We resum logarithms of a moment variable $N$, conjugate to a kinematical 
variable that measures distance from threshold.
We then write the resummed cross section as \cite{NKGS}
\beqa
\sigma(N) =
\exp\left[ \sum_i E_i(N)\right] \;
H\left(\alpha_s\right) \,
\exp \left[\int_{\sqrt{s}}^{{\sqrt{s}}/{\tilde N}}
\frac{d\mu}{\mu} \;
\Gamma_S^{\dagger}(\mu)\right] \;
S \left(\frac{\sqrt{s}}{\tilde N}\right)\;
\exp \left[\int_{\sqrt{s}}^{{\sqrt{s}}/{\tilde N}}
\frac{d\mu}{\mu}\; \Gamma_S(\mu)\right] 
\label{resHS}
\eeqa
where $E_i$ resums universal collinear and soft-gluon emission from the 
incoming partons, $H$ is the hard-scattering function, $S$ is the 
soft-gluon function describing noncollinear soft-gluon emission, and 
$\Gamma_S$ is the soft anomalous dimension - a matrix in color space
and a function of kinematical invariants.
We calculate $\Gamma_S$ in the eikonal approximation, in momentum space and 
using Feynman gauge, from the UV poles of dimensionally-regularized 
eikonal diagrams.

\section{Two-loop soft anomalous dimensions}

Complete two-loop results for $\Gamma_S$ are now available for
the soft (cusp) anomalous dimension for $e^+ e^- \rightarrow t {\bar t}$, 
for $s$-channel single top production, for  
$bg \rightarrow t W^-$ and $bg \rightarrow t H^-$, and 
for $t{\bar t}$ hadroproduction.

\subsection{Soft (cusp) anomalous dimension for 
$e^+ e^- \rightarrow t {\bar t}$}

The soft (cusp) anomalous dimension for $e^+ e^- \rightarrow t {\bar t}$ is
$\Gamma_S=(\alpha_s/\pi) \Gamma_S^{(1)}+(\alpha_s/\pi)^2 \Gamma_S^{(2)}+\cdots
$ with
$\Gamma_S^{(1)}=C_F (\gamma \coth\gamma-1)$, 
in terms of the cusp angle  
$\gamma=\ln[(1+\beta)/(1-\beta)]$ with $\beta=\sqrt{1-4m^2/s}$
and $m$ the top quark mass, 
and \cite{NK2l}
\beqa
\Gamma_S^{(2)}&=&\frac{K}{2} \, \Gamma_S^{(1)}
+C_F C_A \left\{\frac{1}{2}+\frac{\zeta_2}{2}+\frac{\gamma^2}{2}
-\frac{1}{2}\coth^2\gamma\left[\zeta_3-\zeta_2\gamma-\frac{\gamma^3}{3}
-\gamma \, {\rm Li}_2\left(e^{-2\gamma}\right)
-{\rm Li}_3\left(e^{-2\gamma}\right)\right] \right.
\\ && \hspace{25mm} \left.
{}-\frac{1}{2} \coth\gamma\left[\zeta_2+\zeta_2\gamma+\gamma^2
+\frac{\gamma^3}{3}+2\, \gamma \, \ln\left(1-e^{-2\gamma}\right)
-{\rm Li}_2\left(e^{-2\gamma}\right)\right] \right\}
\eeqa
where 
$K=C_A (67/18-\zeta_2)-5n_f/9$.

The cusp anomalous dimension written above is an essential component of 
calculations for other QCD processes, where the color structure gets more 
complicated with more than two colored partons involved in the process.

\subsection{Soft anomalous dimensions for single-top production processes}

Next, we compute two-loop soft anomalous dimensions for 
single top production in the $s$-channel and   
for the associated production of a top quark 
with a $W$ boson or a charged Higgs.

For $s$-channel single top production the one-loop \cite{NKst,NKsch2l} 
and two-loop \cite{NKsch2l} expressions are
\beqa 
\Gamma_{S,\, {\rm top \; s-ch}}^{(1)}&=&C_F \left[\ln\left(\frac{s-m^2}{m\sqrt{s}}\right)
-\frac{1}{2}\right] 
\\
\Gamma_{S,\, {\rm top \; s-ch}}^{(2)}&=&\frac{K}{2} \Gamma_{S,\, {\rm top \, s-ch}}^{(1)}
+C_F C_A \frac{(1-\zeta_3)}{4} \, .  
\eeqa

For associated production of a top quark with a $W^-$,  $bg \rightarrow tW^-$,
the soft anomalous dimension at one-loop \cite{NKst,NKtWH2l} and two-loops 
\cite{NKtWH2l} is
\beqa
\Gamma_{S,\, tW^-}^{(1)}&=&C_F \left[\ln\left(\frac{m^2-t}{m\sqrt{s}}\right)
-\frac{1}{2}\right] +\frac{C_A}{2} \ln\left(\frac{m^2-u}{m^2-t}\right)
\\
\Gamma_{S,\, tW^-}^{(2)}&=&\frac{K}{2} \Gamma_{S,\, tW^-}^{(1)}
+C_F C_A \frac{(1-\zeta_3)}{4} \, .
\eeqa
We find the same analytical results for the related process 
$bg\rightarrow tH^-$ \cite{NKtH,NKtWH2l}.

\subsection{Soft anomalous dimension matrices for $t{\bar t}$ production}

For top-antitop production in hadron colliders, the 
soft anomalous dimensions were derived at one-loop in \cite{NKGS} and  
have been studied in various approaches at two-loops in 
\cite{NK2l,MSS,BNm,BFS,FNPY,NKnnllttb}. 
For $q{\bar q} \rightarrow t{\bar t}$ the soft anomalous dimension 
is a $2\times 2$ matrix while that 
for $gg \rightarrow t{\bar t}$ is a $3\times 3$ matrix.

The soft anomalous dimension matrix for $q{\bar q} \rightarrow t{\bar t}$ is 
\beqa
\Gamma_{S\, q{\bar q}}=\left[\begin{array}{cc}
\Gamma_{q{\bar q} \, 11} & \Gamma_{q{\bar q} \, 12} \\
\Gamma_{q{\bar q} \, 21} & \Gamma_{q{\bar q} \, 22}
\end{array}
\right] 
\eeqa
At one loop, in a color basis
of singlet and octet exchange in the $s$ channel, we find 
\cite{NKGS,NKnnllttb}
\beqa
&& \Gamma_{q{\bar q} \,11}^{(1)}=-C_F \, [L_{\beta}+1],
\hspace{15mm}
\Gamma_{q{\bar q} \,12}^{(1)}=
\frac{C_F}{C_A} \ln\left(\frac{u_1}{t_1}\right),
\hspace{15mm}
\Gamma_{q{\bar q} \,21}^{(1)}=
2\ln\left(\frac{u_1}{t_1}\right),
\\ &&
\Gamma_{q{\bar q} \,22}^{(1)}=C_F
\left[4\ln\left(\frac{u_1}{t_1}\right)
-L_{\beta}-1\right]
+\frac{C_A}{2}\left[-3\ln\left(\frac{u_1}{t_1}\right)
+\ln\left(\frac{t_1u_1}{s m^2}\right)+L_{\beta}\right],
\eeqa
where 
$L_{\beta}=[(1+\beta^2)/(2\beta)] \ln[(1-\beta)/(1+\beta)]$
with $\beta=\sqrt{1-4m^2/s}$.

In presenting our two-loop results, we use the short-hand notation 
$M_{\beta}$ to denote the terms inside the curly brackets in our previous 
expression for the cusp anomalous dimension.
Then at two loops we find \cite{NKnnllttb}
\beqa
&& \Gamma_{q{\bar q} \,11}^{(2)}=\frac{K}{2} \Gamma_{q{\bar q} \,11}^{(1)}
+C_F C_A \, M_{\beta}, \hspace{20mm}
\Gamma_{q{\bar q} \,12}^{(2)}=
\frac{K}{2} \Gamma_{q{\bar q} \,12}^{(1)} -\frac{C_F}{2} 
N_{\beta} \ln\left(\frac{u_1}{t_1}\right),
\\ &&
\Gamma_{q{\bar q} \,21}^{(2)}=
\frac{K}{2}  \Gamma_{q{\bar q} \,21}^{(1)} +C_A N_{\beta} \ln\left(\frac{u_1}{t_1}\right), \hspace{15mm} 
\Gamma_{q{\bar q} \,22}^{(2)}=
\frac{K}{2} \Gamma_{q{\bar q} \,22}^{(1)}
+C_A\left(C_F-\frac{C_A}{2}\right) \, M_{\beta}, 
\eeqa
with $N_{\beta}$ a subset of the terms of $M_{\beta}$.

For the $g(p_a) +g(p_b) \rightarrow t(p_1)+{\bar t}(p_2)$ channel 
we choose the color basis:
$c_1=\delta^{ab}\,\delta_{12}$, $c_2=d^{abc}\,T^c_{12}$,
$c_3=i f^{abc}\,T^c_{12}$. 
The soft anomalous dimension matrix for $gg \rightarrow t{\bar t}$ 
in this basis is 
\beqa
\Gamma_{S \, gg}=\left[\begin{array}{ccc}
\Gamma_{gg\, 11} & 0 & \Gamma_{gg\,13} \vspace{2mm} \\
0 & \Gamma_{gg\,22} & \Gamma_{gg\,23} \vspace{2mm} \\
\Gamma_{gg\,31} & \Gamma_{gg\,32} & \Gamma_{gg\,22}
\end{array}
\right] \, .
\label{matrixggtt}
\eeqa
At one loop we have \cite{NKGS,NKnnllttb}
\beqa
&& \Gamma_{gg\, 11}^{(1)}=-C_F [L_{\beta}+1],
\hspace{5mm}
\Gamma_{gg\, 22}^{(1)}=-C_F [L_{\beta}+1] 
+\frac{C_A}{2}\left[\ln\left(\frac{t_1 u_1}{m^2 s}\right)+L_{\beta}\right],
\hspace{5mm}
\Gamma_{gg\, 23}^{(1)}=\frac{C_A}{2} \ln\left(\frac{u_1}{t_1}\right),
\\
&& \Gamma_{gg\, 13}^{(1)}= \ln\left(\frac{u_1}{t_1}\right),
\hspace{15mm}
\Gamma_{gg\, 31}^{(1)}=2 \, \ln\left(\frac{u_1}{t_1}\right),
\hspace{15mm}
\Gamma_{gg\, 32}^{(1)}=\frac{N_c^2-4}{2N_c} \ln\left(\frac{u_1}{t_1}\right).
\label{Gamma1ggtt}
\eeqa
At two loops we find \cite{NKnnllttb}
\beqa
&& \Gamma_{gg\, 11}^{(2)}= \frac{K}{2} \Gamma_{gg \,11}^{(1)}
+C_F C_A \, M_{\beta}, \hspace{5mm} 
\Gamma_{gg\, 22}^{(2)}= \frac{K}{2} \Gamma_{gg \,22}^{(1)}
+C_A \left(C_F-\frac{C_A}{2}\right) \, M_{\beta},
\hspace{5mm} 
\Gamma_{gg\, 23}^{(2)}=\frac{K}{2} \Gamma_{gg \,23}^{(1)},
\\
&& \Gamma_{gg\, 13}^{(2)}=\frac{K}{2} \Gamma_{gg \,13}^{(1)} 
-\frac{C_A}{2} N_{\beta} \ln\left(\frac{u_1}{t_1}\right),
\hspace{5mm}
\Gamma_{gg\, 31}^{(2)}=\frac{K}{2} \Gamma_{gg \,31}^{(1)} 
+C_A N_{\beta} \ln\left(\frac{u_1}{t_1}\right),
\hspace{5mm} 
\Gamma_{gg\, 32}^{(2)}=\frac{K}{2} \Gamma_{gg \,32}^{(1)}. 
\label{Gamma2ggtt}
\eeqa

\end{document}